\documentclass[aps,prl,twocolumn,showpacs,groupedaddress]{revtex4-1}

\usepackage{graphicx}  
\usepackage{epstopdf}
\usepackage{dcolumn}   
\usepackage{bm}        
\usepackage{amssymb}   

\begin{document}

\title{Resolving Spin-Orbit and Hyperfine Mediated Electric Dipole Spin
Resonance in a Quantum Dot}

\author{M. Shafiei$^{1}$, K. C. Nowack$^{1}$, C. Reichl$^{2}$, W.
Wegscheider$^{2}$}
\author{L. M. K. Vandersypen$^{1}$}
\affiliation{$^{1}$Kavli Institute of Nanoscience, Delft University of
Technology, PO Box 5046, 2600 GA Delft, The Netherlands}
\affiliation{$^{2}$Solid State Physics Laboratory, ETH Zurich,
Schafmattstrasse 16, 8093 Zurich, Switzerland}

\date{\today}

\begin{abstract}
We investigate the electric manipulation of a single electron spin in a single gate-defined quantum dot. We observe that so-far neglected differences between the hyperfine and spin-orbit mediated electric dipole spin resonance conditions have important consequences at high magnetic fields. In experiments using adiabatic rapid passage to invert the electron spin, we observe an unusually wide and asymmetric response as a function of magnetic field. Simulations support the interpretation of the lineshape in terms of four different resonance conditions. These findings may lead to isotope-selective control of dynamic nuclear polarization in quantum dots.\end{abstract}
 
\pacs{76.30.-v,73.63.Kv,03.67.-a}

\maketitle

Manipulation of electron spins is an essential
tool for applications in spin electronics
(spintronics) \cite{Zutic2004,Awschalom2002}. In the limit of single-electron spin manipulation, applications in solid-state quantum computation arise, where the electron spin serves as a two-level system (qubit) \cite{Loss1998}. Conventionally the manipulation of electron spins makes use of electron spin
resonance (ESR) whereby an alternating magnetic field is applied with a frequency equal to 
the precession frequency of the electron spin \cite{Poole1983}. 
In semiconductor quantum dots, single electron spin manipulation by ESR has been realized by applying a large localized
alternating magnetic field at low temperature, which is challenging \cite{Koppens2006}. In comparison, it is much easier to create and localize an oscillating electric field.
In a semiconductor environment, electric fields can couple to electron spins, and electron spin transitions can be induced through electric-dipole spin resonance
(EDSR) \cite{Bell1962,McCombe1967,Kato2003,Sheka1991}.
Recently, EDSR has been measured on single electron spins in quantum dots 
\cite{Nowack2007,Laird2007,Pioro-Ladriere2008,Nadj-Perge2010}.

The coupling of electric fields to the electron spin can be mediated in several ways: a transverse magnetic field gradient \cite{Pioro-Ladriere2008,Tokura2006}, exchange with magnetic impurities \cite{Khazan1993}, a position dependent $g$-tensor \cite{Kato2003}, spin-orbit coupling \cite{Sheka1991,Golovach2006,Nowack2007,Levitov2003,Walls2007}, and hyperfine interaction with nuclear spins
\cite{Laird2007}. A unified theoretical description is given in \cite{Rashba2008}.
In all these experiments and theoretical discussions, any differences
in the resonance frequencies associated with the different driving mechanisms have been
neglected.

Here we show that at high magnetic fields there is a clearly observable shift in the resonance condition between spin-orbit mediated (SO-EDSR) and hyperfine mediated EDSR (HF-EDSR). In these experiments, we introduce adiabatic rapid passage as a robust technique to invert the electron spin in quantum dots using fast frequency chirps, since surprisingly no EDSR response is obtained when using fixed-frequency excitation. Furthermore, by modeling the EDSR response, we get a deeper understanding of the interplay between spin-orbit and hyperfine mediated driving.

\begin{figure}
\includegraphics[width=8.5cm] {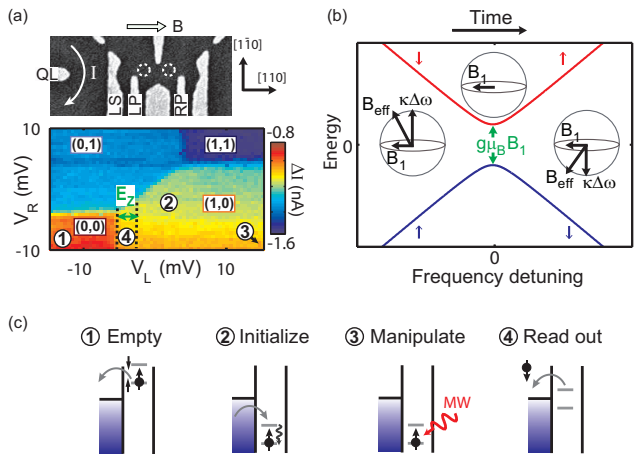}
\caption{\label{fig:fig1} (a) Scanning electron micrograph of a device similar
to the one measured (top) and QPC current (a background plane has been
subtracted) as function of LP and RP gate voltages (bottom).
$V_{\text{L}}=1.2V_{\text{LP}}+0.6V_{\text{RP}}$ and $V_{\text{R}}=1.4V_{\text{LP}}+0.5V_{\text{RP}}$. $(n,m)$ indicate the number of electrons in the left and right dot, respectively. 
(b) Schematic explanation of adiabatic rapid passage (see text). Blue and red solid lines show the electron
spin eigenenergies in the presence of an ac excitation transverse to the static magnetic field as a function of detuning from resonance. In the experiment, the excitation frequency was chirped in the direction of the time
arrow. The insets show the effective field acting on the spin for three values of detuning ($\kappa=\hbar/g \mu_{\text{B}}$, with $\mu_B$ the Bohr magneton and $g$ the electron $g$-factor). (c) Electrochemical potential diagrams during the pulse cycle (see text). At the
read-out stage, the electron tunnels out if and only if it is spin down, in which case a step in the QPC current results.}
\end{figure}

The measurements are performed on a single quantum dot laterally defined in a GaAs/(Al,Ga)As heterostructure. A quantum point contact (QPC) is placed close to the quantum dot and used as a charge detector \cite{Field1993}. 
Figure \ref{fig:fig1}(a) shows a device image and the charge stability diagram in the region of interest. The electron temperature is about 250 mK.
An in-plane magnetic field is applied along the $[110]$ crystallographic axis to split the spin states. We control the electrochemical potential of the dot by applying voltage pulses on gates LP and RP, which are fitted with bias tees.

The measurement scheme is as follows (Fig. \ref{fig:fig1}(c)):
first we empty the dot, inject an electron in the left
quantum dot by pulsing from the (0,0) to the (1,0) charge state, and initialize to spin-up by waiting longer than the spin relaxation time, close to the (1,0)-(0,1) charge transition where relaxation is fastest \cite{relaxcomment} ($\sim 6$ ms altogether). Then we apply a microwave burst to gate LS to manipulate the electron spin (manipulation step, 100-500 $\mu$s). The manipulation step is done deep in Coulomb blockade in order to minimize the effect of photon-assisted tunneling and avoid thermal excitation to/from the reservoirs. Finally we pulse into the read-out position and measure the electron spin state using energy selective spin-to-charge conversion \cite{Elzerman2004,Nowack2011} (1.5 ms, single shot read-out fidelities were $\sim 95\%$ and $\sim 80\%$ for the detection of spin-up and spin-down states, respectively). We end with a 2.5 ms compensation stage that makes the pulse amplitude average to zero, so no offsets are induced by the bias-tee. Sufficient repetition of this cycle (100-2000 repetitions) yields the spin-down probability at the end of the manipulation stage.

The microwave (MW) burst is chosen to be long
enough ($\tau_{\text{MW}}=400$ $\mu$s) compared to the Rabi decay time ($\sim1$ $\mu$s \cite{Koppens2006}) to create a mixture of
spin-up and spin-down states. The applied microwave power is maximized by operating just below the onset of photon-assisted tunneling. 
The magnitude of the electric field is estimated to be of the same order of magnitude as in previous
EDSR measurements in GaAs quantum dots \cite{Laird2007,Nowack2007}.
To find the electron spin resonance position, we apply such pulses at different
static magnetic fields.  

\begin{figure}
\includegraphics[width=8.5cm] {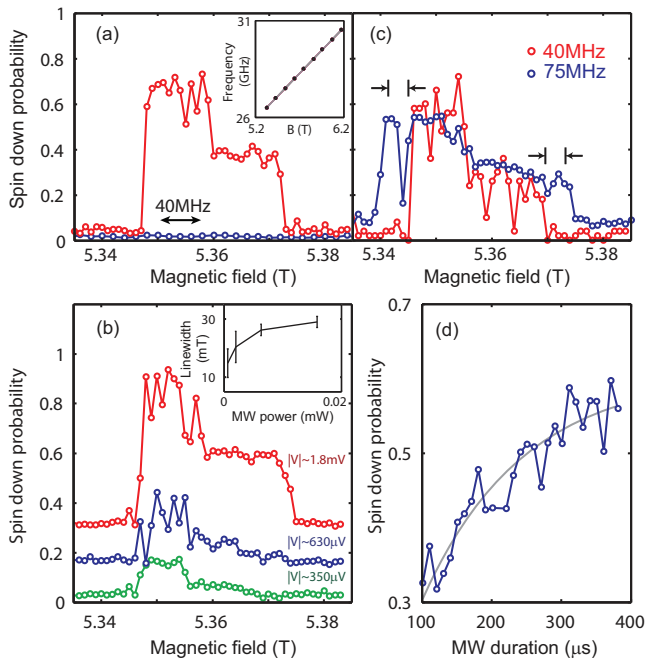}
\caption{\label{fig:fig2}(a) Measured spin-down probability as a function of magnetic field for a $400 \mu$s MW burst at $f_{\text{MW}}=26.5$
GHz (blue) and for a $400 \mu$s frequency chirped MW burst with a frequency modulation (FM) depth of $40$ MHz centered at
26.5 GHz (red). The double arrow shows the magnetic field span corresponding to $40$
MHz. Inset: EDSR resonance frequency as a
function of magnetic field, along with a linear fit (gray curve), giving an electron g-factor of $-0.339\pm0.003$. As the resonance position we took half a FM depth above the left flank of the response. (b) Similar to (a), for three different driving amplitudes. The amplitudes are estimated considering the attenuations in the MW line (traces are offset for clarity). Inset: The linewidth saturates as the power increases. 
(c) Similar to (a) for two different FM depths, both at a chirp rate of $150$ MHz/ms. Arrows show the expected difference in the linewidth. (d) Measured spin down probability at
$\text{B}=5.349$ T as a function of MW burst duration (FM depth $75$ MHz).
The gray line is an exponential fit to the data. All
measurements in (a,c,d) are taken with the same MW power as the red trace in (b).}
\end{figure}

Surprisingly, under fixed-frequency excitation, no EDSR response is observed, regardless of the magnetic field sweep direction (Fig. \ref{fig:fig2}(a)). We used magnetic field steps down to 1 mT, well below the 5-10 mT linewidth typically observed in GaAs quantum dots due the (1-2 mT rms) distribution of nuclear fields \cite{Koppens2006,Nowack2007, Koppens2007}. 
Based on the geometry we expect the alternating field to be mostly along
$[110]$, so the Dresselhaus and Rashba contributions to the spin-orbit interaction work against each other \cite{Golovach2006,SOComment}. If they largely cancel out, SO-EDSR is very weak, but we would still expect a signal due to hyperfine mediated EDSR \cite{Laird2007}. The resonance condition could shift due to dynamic nuclear polarization (DNP), but this is expected to occur only for one sweep direction of the magnetic field \cite{Danon2008} and is thus not likely the explanation. We return to the absence of a fixed-frequency  response near the end of this Letter.

Remarkably, a very strong and clear single-spin EDSR response was obtained when the microwave bursts were frequency chirped (red trace in Fig. \ref{fig:fig2}(a)): when the excitation frequency passes through the resonance frequency under the right conditions, the electron spin is inverted in a process called adiabatic rapid passage \cite{Abragam1983}. This process can be understood as follows (see Fig. \ref{fig:fig1}(b)). The microwave excitation produces an oscillating (effective) magnetic field in the equatorial plane of the Bloch sphere. In the rotating frame synchronized with the instantaneous excitation frequency, the driving field lies along a fixed axis ($B_1$ in the insets of Fig. \ref{fig:fig1}(b)). When the microwave frequency is detuned from the spin Larmor frequency by $\Delta\omega$, the electron is subject to an additional effective magnetic field with magnitude $\hbar\Delta\omega/g\mu_{\text{B}}$ perpendicular to the equatorial plane of the Bloch sphere (left and right inset). When sweeping the MW frequency from far below to far above the resonance frequency, the total effective field, $B_{\text{eff}}$, will get inverted. The electron spin will track the total effective field and get inverted as well, provided the frequency sweep rate is much slower than the Rabi frequency, $\Omega$, squared (adiabaticity condition) and spin coherence is preserved during the inversion \cite{Shevchenko2010}. For large MW chirp ranges, adiabatic inversion is insensitive to the exact value of the resonance frequency and is thus robust to slow fluctuations of the resonance position.

\begin{figure}  
\includegraphics[width=8.5cm] {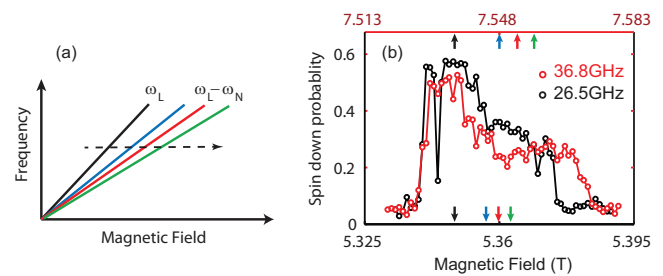}
\caption{\label{fig:fig3} (a) Schematic representation of the resonance
condition of SO-EDSR, $\omega_L$ (black), and HF-EDSR,
$\omega_L-\omega_N$, for $^{75}As$ (blue), $^{69}Ga$ (red)
and $^{71}Ga$ (green) nuclei. The separations are not to scale. The dashed-arrow
shows the direction of the magnetic field sweep in the measurements. (b) Measured spin-down probability versus magnetic field for a 75 MHz chirp at $26.5$ GHz
(black) and $36.8$ GHz (red). Both measurements have been done at high MW power, where the linewidth is saturated. Arrows show the relative positions of spin-orbit and hyperfine mediated EDSR resonances (same color coding as in (a)). }
\end{figure}  

The response in Fig. \ref{fig:fig2}(a) is very strong but the lineshape is unexpected. For adiabatic rapid passage through a single resonance and
in the presence of random nuclear fluctuations, the
lineshape is expected to be symmetric and the convolution
of a boxcar function with width equal to the FM depth and
a Gaussian distribution associated with the nuclear spin fluctuations.  
The observed lineshape, in contrast, is asymmetric and
the linewidth ($\sim26\pm4$ mT) is much larger than both the FM depth
(corresponding to $8.4$ mT) and the random nuclear field. The asymmetry and width are reproducible, with variations between repeated measurements with identical settings about as large as the variations between the red traces of Figs. \ref{fig:fig2}(a-c). The variations are due to a statistical error in the measured probability and from sampling of the nuclear field distribution (see \cite{supplementary} for more details).
The asymmetric lineshape is reminiscent of lineshapes which have been identified as the result of DNP  \cite{Vink2009}, but unlike those measurements did not show
any hysteretic behavior in the frequency or field sweep direction. 

The behavior of the EDSR peaks as a function of MW power is shown in Fig.
\ref{fig:fig2}(b). For very low MW powers, the width of the EDSR signal is very sensitive to the MW power while
for high MW powers the linewidth saturates (Fig. \ref{fig:fig2}(b) inset).
Furthermore, the width of the response increases with increasing FM depth
(Fig. \ref{fig:fig2}(c); see \cite{supplementary} for additional data using smaller FM depth). Finally, when reducing the burst time 
while keeping the modulation depth constant, the spin flip probability is gradually reduced as well (Fig. \ref{fig:fig2}(d)). This is consistent with a transition from adiabatic passage to diabatic passage upon increasing the chirp rate. From the time constant of the exponential fit to the data and the chirp rate, we can extract a Rabi frequency of $\approx$ 0.2 MHz using the Landau-Zener model for transition probabilities \cite{Shevchenko2010}. Measurements with a microwave chirp back and forth, corresponding to two consecutive adiabatic passages, were also consistent with this model (for details see \cite{supplementary}).

Understanding the lineshape requires consideration of the
different EDSR mechanisms, here spin-orbit and hyperfine mediated EDSR.
The SO-EDSR resonance frequency is equal to the electron spin
Larmor frequency ($\omega_L=g\mu_BB_{ext}/\hbar$ with $B_{ext}$ the external magnetic field). In contrast, HF-EDSR involves direct electron-nuclear spin flip-flops, so the resonance frequency is smaller than that of
SO-EDSR by the nuclear spin Larmor
frequency, $\omega_N=g_N\mu_NB_{ext}/\hbar$, where $g_N$ and $\mu_N$ are the nuclear g-factor and magneton, respectively (or for a given excitation frequency, the resonant field is higher, as in Fig. \ref{fig:fig3}(b)). For the three nuclear species in GaAs, $^{75}$As, $^{69}$Ga and
$^{71}$Ga, the shift in the resonance conditions amounts to $7.318$ MHz/T, $10.24$ MHz/T and $13.02$ MHz/T, respectively,
giving rise to a total of four resonance conditions (Fig. \ref{fig:fig3}(a)). These shifts are usually neglected, but are important at high field, see the two sets of vertical arrows in Fig. \ref{fig:fig3}(b). The total linewidth observed at 26.5 GHz covers the range of the four resonance conditions plus twice the FM depth, which is a first indication that all resonances play a role. In line with this interpretation, we see in Fig. \ref{fig:fig3}(b) that the response at a $36.8$ GHz center frequency is correspondingly broader than that at $26.5$ GHz.

In the measurements, we see that the signal is systematically higher in the SO-EDSR field range than in the HF-EDSR field range. A first possible explanation could be that the adiabaticity condition is here better satisfied for SO-EDSR than for HF-EDSR. A second possibility is that the separation between the three hyperfine-mediated resonances is so small compared to the Rabi frequencies that neighbouring resonances affect the spin dynamics simultaneously, spoiling spin inversion.

To clarify this issue, we numerically simulate the response to the chirped MW bursts. For simplicity, in the simulation four excitation frequencies are chirped through a single resonance frequency $\omega_L$, rather than sweeping one excitation through four resonances. Then the Hamiltonian in the laboratory frame is given by 
 \begin{eqnarray*}
-\frac{\hbar}{2} \omega_L \sigma_z -
\hbar \bigg[ \Omega_{so} \cos(\omega t)+
\sum_N\Omega_{hf}^{N} \cos((\omega+\omega_N) t)\bigg]\sigma_x
\end{eqnarray*}
where $\Omega_{so}$ and $\Omega_{hf}^N$ are the spin-orbit mediated and hyperfine
mediated Rabi frequencies, with nuclear species $N=^{75}$As, $^{69}$Ga, $^{71}$Ga; 
$\sigma_{x,z}$ are the $x$ and $z$
Pauli matrices. In the simulations, we approximate the effect of the time dependence of the nuclear field during the chirp using a phase damping operator \cite{PDComment}, using $T_2 = 100 \mu$s for the coherence time (the simulation results were insensitive to the value of $T_2$ in the range $\sim100-500$ $\mu$s). Additionally, we account for the random nuclear field at the start of every cycle by convoluting the response with a normal distribution function with standard deviation 0.5 mT \cite{SigmaComment}.

\begin{figure} 
\includegraphics[width=8.5cm] {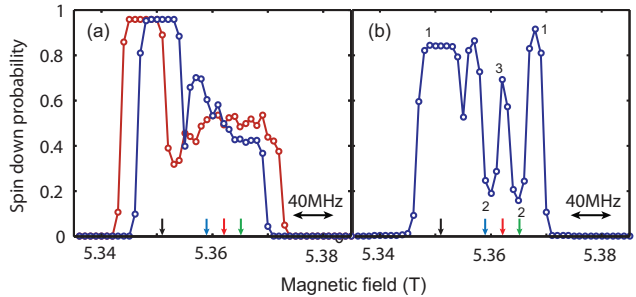}
\caption{\label{fig:fig4} Simulated spin-down probability (assuming perfect measurement fidelity) versus magnetic field for a 500 $\mu$s, 40 MHz chirp (blue) and a 937 $\mu$s, 75 MHz chirp (red) for (a) 
$\Omega_{so}=1.25$ MHz, $\Omega_{hf}^{^{75}As}=0.63$ MHz and 
(b) $\Omega_{so}=6.28$ MHz, $\Omega_{hf}^{^{75}As}=5.02$ MHz. The numbers in panel (b) express the number
of resonances covered by the chirp. Arrows show spin-orbit
and hyperfine mediated EDSR resonance positions (color coding as in Fig. \ref{fig:fig3}). Double arrows show the magnetic field range corresponding to the FM depth.
}
\end{figure}

The spin-orbit mediated Rabi frequency used in the simulation was that extracted in Fig. \ref{fig:fig2}(d), $\Omega_{so} = 2\pi\times0.2$ MHz (that data is taken at a magnetic field where the chirp range does not cover the hyperfine-mediated resonances). We do not have an independent measurement of $\Omega_{hf}^N$, but we do know the ratio between the three $\Omega_{hf}^N$ \cite{hfRabicomment}. We have extensively explored the response for various values for $\Omega_{hf}^N$ in the simulation and found an asymmetric response as observed in the experiment when $\Omega_{so} > \Omega_{hf}$, with $0.6$ MHz $<\Omega_{so}<5$ MHz and $0.3$ MHz $<\Omega_{hf}^{As}<0.8$ MHz. An example lineshape is shown in Fig. \ref{fig:fig4}(a). The adiabaticity condition ($\Omega^2> \frac{2}{\pi} d\omega/dt$) is then better satisfied and spin inversion is more complete for spin-orbit mediated driving than for hyperfine mediated driving. This explains the observed asymmetric lineshape.
Similar to the measurement of Fig. 2(c), increasing the chirp range to 75 MHz while keeping the FM rate constant broadens the line (red trace in Fig. 4(a)) since the resonances can now be reached from a larger field range. The steepness of flanks remains the same, since it mostly depends on the nuclear field distribution.

The adiabaticity condition is well satisfied for all four resonances for Rabi frequencies 5-10 times higher than in the present experiment, which should still be achievable \cite{Laird2007,Nowack2007,hfRabicomment}. Then each of the resonances by itself is capable of inverting the spin during a frequency chirp, and a striking response is obtained (Fig. \ref{fig:fig4}(b)). The signal is either high or low depending on whether the chirp range covers an odd respectively even number of resonances. This corresponds to an odd respectively even number of spin inversions during the chirp, as explicitly verified in the simulation by looking at the spin state as a function of time throughout the chirp. This behavior requires Rabi frequencies smaller than the separation between adjacent resonances, so that at most one resonance acts on the spin at any given time.

We now return to the absence of a fixed-frequency response in Fig. \ref{fig:fig2}(a). From Fig. \ref{fig:fig2}(d) we estimate that the driving field from spin-orbit interaction is at most $\sim 0.05$ mT. Based on comparison of the measured and simulated lineshape, HF-EDSR may be even weaker. 
The driving field is thus much smaller than the 1-2 mT standard deviation of the nuclear field distribution.
 Given the $\sim 1$ s autocorrelation time of the nuclear field \cite{Reilly2008}, it is possible that the resonance condition fluctuated too far within the measurement time that the signal was missed (the shortest measurement time was also about 1 s, for 100 cycles). The frequency chirp, in contrast, always passes through the resonance, regardless of its exact position.

In conclusion, we have shown that at high magnetic fields it is
possible to separate spin-orbit and hyperfine mediated EDSR by their different resonance conditions. These differences could be exploited for enhanced control of DNP processes, including selective control of the three nuclear spin species. Furthermore, adiabatic
rapid passage is a robust technique for EDSR spectroscopy and spin inversion in III-V quantum dots due to its
robustness to a randomly fluctuating resonance position.

\begin{acknowledgments}
We acknowledge useful discussions with F. Braakman, E. Laird, L. Schreiber, technical support by R. Schouten, and financial support by the Intelligence Advanced Research Projects Activity, European Research Council, Dutch Foundation for Fundamental Research on Matter and Swiss National Science Foundation.
\end{acknowledgments}


\begin{thebibliography}{36}
\bibitem [{SOC()}]{SOComment}%
  \BibitemOpen
  \href@noop {} {}\bibinfo {note} {Band structure calculations indicate
  opposite signs for the Dresselhaus and Rashba spin-orbit coupling
  coefficients in our heterostructure.}\BibitemShut {Stop}%
\bibitem [{PDC()}]{PDComment}%
  \BibitemOpen
  \href@noop {} {}\bibinfo {note} {The electron spin density matrix $\rho$ is
  transformed to $\sum_{i=0,1} E_i\rho E_i^\dagger$ where $E_0=\sqrt{p}I$,
  $E_1=\sqrt{1-p}\sigma_z$, $p=\frac{1+exp(-t/T_2)}{2}$ \cite{nielsenchuang}.
  Since these are single-shot measurements, it is $T_2$ and not $T_2^*$ that
  governs decoherence during each cycle.}\BibitemShut {Stop}%
\bibitem [{Sig()}]{SigmaComment}%
  \BibitemOpen
  \href@noop {} {}\bibinfo {note} {Nuclear field fluctuations of 0.5 mT gave
  better agreement with the data than the usual 1-2 mT, presumably since only
  part of the distribution is sampled within the measurement time.}\BibitemShut
  {Stop}%
\end{thebibliography}
\end{document}